\documentclass[12pt,preprint]{aastex}

    \def\beq{\begin{equation} }
    \def\eeq{\end{equation} }
    \def\spose#1{\hbox to 0pt{#1\hss}}
    \def\ltsim{\mathrel{\spose{\lower.5ex\hbox{$\mathchar"218$}}
     \raise.4ex\hbox{$\mathchar"13C$}}}

\def\spose#1{\hbox to 0pt{#1\hss}}
\def\lta{\mathrel{\spose{\lower 3pt\hbox{$\mathchar"218$}}
        \raise 2.0pt\hbox{$\mathchar"13C$}}}
\def\gta{\mathrel{\spose{\lower 3pt\hbox{$\mathchar"218$}}
        \raise 2.0pt\hbox{$\mathchar"13E$}}}
\shorttitle{All quiet in Globular Clusters}
\shortauthors{Dobrotka, Lasota \& Menou}

\begin{document}

\title{All quiet in Globular Clusters}

\author{Andrej Dobrotka}
\affil{Slovak University of Technology in Bratislava, Department of
 Physics, Faculty of Material Sciences and Technology, Paulinska 16,
 Trnava 91724, Slovakia\\
\& Institut d'Astrophysique de Paris, UMR 7095 CNRS, Universit\'e
Pierre \& Marie Curie, 98bis Bd. Arago, 75014 Paris, France}
\author{Jean-Pierre Lasota}
\affil{Institut d'Astrophysique de Paris, UMR 7095 CNRS, Universit\'e
Pierre \& Marie Curie, 98bis Bd. Arago, 75014 Paris, France}
\author{Kristen Menou}
\affil{Department of Astronomy, Columbia University, 550 West 120th Street,
New York, NY 10027, USA}

\begin{abstract}

Cataclysmic Variables (CVs) should be present in large numbers in
Globular Clusters (GCs).  Numerous low-luminosity X-ray sources
identified over the past few years as candidate CVs in GCs support
this notion. Yet, very few "cataclysms," the characteristic feature of
this class of objects in the field, have been observed in GCs. We
address this discrepancy here, within the framework of the standard
Disk Instability Model for CV outbursts. We argue that the paucity of
outbursts in GCs is probably not a direct consequence of the donors'
low metallicities.  We present diagnostics based on outburst
properties allowing tests of the hypothesis that rare cataclysms are
entirely due to lower mass transfer rates in GCs relative to the
field, and we argue against this explanation.  Instead, we propose
that a combination of low mass transfer rates ($\gta
10^{14-15}$~g~s$^{-1}$) and moderately strong white dwarf magnetic
moments ($\gta 10^{30}$~G~cm$^3$) stabilize CV disks in GCs and thus
prevent most of them from experiencing frequent outbursts. If it is
so, rare cataclysms in GCs would signal important evolutionary
differences between field and cluster CVs.

\end{abstract}


\section{Introduction}

X-ray observations of globular clusters have uncovered many compact
binaries \citep[for a review see][]{VW04}. Recent years brought
considerable progress in our understanding of these sources, thanks to
observations with the Chandra X-ray Observatory and further
identifications with the Hubble Space Telescope
\citep{Ed03a,Ed03b,Heinke03,Hagg04}. Among the weak ($L_X \lta 10^{34}
\rm erg\ s^{-1}$) X-ray sources discovered, one identifies cataclysmic
variable stars (CVs), quiescent low-mass X-ray binaries (qLMXBs) and
binaries containing stars with active coronae. The presence of a large
number of CVs in globular clusters is expected on theoretical grounds
\citep[e.g.][]{Clarck75,Katz75,distef,ivanova,towbil}.

CVs are close binary systems in which a white dwarf accretes matter
lost by a Roche-lobe filling low-mass stellar companion
\citep[see][]{Warner95}. In CVs of AM-Her type (also called
``polars"), the white-dwarf's magnetic moment is sufficiently strong
to synchronize its spin {with} the binary's orbital rotation and to
preclude the formation of an accretion disk. Polars are soft X-ray
emitters with luminosities $L_X \sim 10^{30} - 10^{32} \rm erg\
s^{-1}$ \citep[e.g.][]{Verbetal97}. In intermediate polars
(IPs),\footnote{We shall restrict the use of the term ``IP" to
(non-synchronous) systems which have detectable or directly inferable
magnetic fields. The lowest detected field in an IP is $7 \times 10^6$
G \citep{WW00}} magnetic moments are weaker and the white-dwarf's
rotation is not synchronous with the orbital motion.  These systems
may or may not possess an accretion disk depending on their specific
parameters \citep[see e.g.][]{KL91,Hell02}. When the disk is present,
it is truncated by the white-dwarf's magnetic field. IPs are the
strongest X-ray emitters among CVs, with luminosities $L_X \sim
10^{31} - 10^{33} \rm erg\ s^{-1}$ \citep[][]{Verbetal97}. In high
optical-luminosity persistent CVs of the nova-like type and in
erupting dwarf-nova systems, the presence of accretion disks is well
established, but constraints on the weaker white dwarf magnetic
moments are generally poor {\citep[see e.g.][]{Warner04}}. ``Normal
outbursts'' in dwarf novae have amplitudes of 2-5 mag and last 2-20
days. Their recurrence times are typically from $\sim 10$ days to
months, and up to years in some cases. Some dwarf novae show
``superoutbursts,'' which have brighter amplitudes by $\sim$ 0.7
magnitudes, last ~$\sim 5$ times longer and also have recurrence times
longer than those of normal outbursts. Nova-like CVs and dwarf novae
in outburst are rather weak X-ray emitters, but in their low-state,
quiescent dwarf novae have typical X-ray luminosities $L_X \sim
10^{29} - 10^{32}~ \rm erg\ s^{-1}$
\citep[][]{PR85,Eracetal91,Verbetal97}.

CVs identified in globular clusters have X-ray luminosities very
similar to those of field CVs, even though some differences exist in
the luminosity functions and one finds that {cluster CVs have on
average} higher optical to X-ray luminosity ratios \citep{Ed03b}. One
is tempted to classify candidate CVs in GCs as quiescent dwarf novae,
or perhaps IPs, on the basis of their X-ray luminosities.  Since in
the solar neighborhood, about half of the observed CVs show large
amplitude, recurrent outbursts, the same should be true of CVs in
globular clusters, if the the two populations have similar
characteristics. This expectation is not supported by observations.
For instance, among the 22 confirmed CVs in 47~Tuc only 2 showed
variability that could be interpreted as a dwarf-nova outburst
\citep{Ed03b}. In the dense cluster M15, only one erupting dwarf-nova
was observed \citep{Sharetal04,Tua03}. Other observational studies
have consistently found that dwarf nova outbursts in GCs are rare
\citep[e.g.][]{Bond05,Kal05,Sha05}.

This strange quietness of CVs in globular clusters has been noticed
and commented upon before \citep[e.g.][]{Sharetal96,G99}. One
explanation offered is that most globular cluster CVs are strongly
magnetic: the magnetic field of the white dwarf could suppress the
instability at the origin of the dwarf-nova outbursts by truncating
otherwise unstable regions of the disk \citep[see
e.g.][]{Ed03b}. According to this proposal, the magnetic fields would
be of the IP strength, i.e.  $\gta 10^6$ G. As it turns out, however,
at least 7 field IPs show dwarf-nova outbursts
\citep{HMB97,Ishiokaetal02}. Not all of these systems exhibit bona
fide dwarf-nova outbursts but several have $~3.5$ mag outbursts
lasting 5 days or longer. In addition, \citet{Sharetal05} observed in
NGC 6397 two (unexpected) dwarf-novae outbursts from CVs classified as
magnetic on the basis of their HeII 4686 emission lines. Even though
most known IPs (more than 30 have been identified) do not indeed show
dwarf-nova outburst, it is clear that both in the field and in
globular clusters, some magnetic CVs still undergo dwarf-nova
outbursts. This indicates that strong magnetic fields alone may not be
able to explain the paucity of outbursts in GCs.

As an alternative, it has also been proposed that globular cluster CVs
accrete at much lower rates than their counterparts in the field
\citep[e.g.][]{Ed03b}. This would lead to less frequent outbursts and
could in principle explain rare cataclysms in GCs. However, to this
day, there has been no detailed, quantitative investigation of this
proposal.  Implicitly, the low mass transfer rate hypothesis invokes
the disk model instability (DIM) to describe (the lack of) dwarf-nova
outbursts \citep[for a recent review of the model see][]{NAR}. Here,
we use the DIM to determine whether indeed low mass transfer rates
alone could explain the very infrequent outbursts of candidate CVs in
GCs. We argue that, given the observed X-ray luminosities of these
systems, it is not a satisfactory explanation. Instead, we propose a
solution to the rare cataclysm problem that invokes a combination of
low mass transfer rates and moderately strong white dwarf magnetic
moments to stabilize the disks of candidate CVs in GCs against
outbursts. We determine quantitatively the conditions which must be
satisfied by the mass transfer rates and the magnetic moments in these
systems to allow globular clusters to be quiet systems of stars,
rather than being the sites of incessant dwarf-nova outbursts.

In \S2, we explain why, on the basis of DIM predictions and observed
X-ray luminosities in GCs, one would naively expect candidate CVs in
GCs to be subject to dwarf nova outbursts. In \S3, we argue that, even
though the low metallicity of the gas accreted in GCs affects the
ionization level of the disk, it is unlikely to be the explanation
behind rare cataclysms in these systems.  In \S4, we investigate
whether it is possible for these candidate CVs to be subject to much
less frequent outbursts than field CVs by having, on average, much
lower mass transfer rates than field CVs. We conclude that it is not
easily the case and explore in \S5 an alternative possibility
involving a stabilization of the disks at low mass transfer rates by
truncation due to moderately strong white dwarf magnetic fields. This
appears to be the most satisfactory explanation for rare cataclysms in
GCs. In \S5, we comment on additional consequences and possible
extensions of our work.

\section{Disk Stability and X-ray Luminosities}
\label{stable}

In the disk instability model (DIM), outbursts of a steadily-fed, thin
accretion disk are caused by a thermal-viscous instability in regions
of partial gas ionization (leading to large opacity variations). There
exist a range of accretion rates for which the instability operates,
determined by the existence of partial ionization regions.  According
to the model, there are three possible global regimes of accretion for
a disk: a hot-stable regime at high accretion rates, an unstable
regime at intermediate accretion rates and a cold-stable regime at low
accretion rates. The critical rates determining locally whether or not
the disk is partially ionized and unstable are strong functions of
radius:
\begin{eqnarray}
\dot{M}_{\rm A} &=& 4.0 \times  10^{15} ~ \alpha^{-0.004} \left(
{M_1 \over \rm M_\odot} \right)^{-0.88} \left( {r \over 10^{10} \
\rm cm} \right)^{2.65}
~\rm g~s^{-1}
\label{mdmax1}\\
\dot{M}_{\rm B} &=& 9.5 \times 10^{15} ~ \alpha^{0.01} \left( {M_1
\over \rm M_\odot} \right)^{-0.89} \left( {r \over 10^{10} \ \rm cm}
\right)^{2.68} ~\rm g~s^{-1},
\label{mdmin1}
\end{eqnarray}
where $M_1$ is the white-dwarf's mass, $\alpha$ is the usual viscosity
parameter, $r$ the radial distance from the center of mass and these
specific criteria refer to a disk with solar abundance material. At
each radius $r$, $\dot{M}_{\rm A}$ is the critical accretion rate
below which the disk is cold/neutral and stable and $\dot{M}_{\rm B}$
is the critical value above which the disk is hot/ionized and
stable. Note that, in the DIM, it is necessary to postulate different
values of $\alpha$ for the disk in the hot state and in the cold state
($\alpha_{\rm hot} \sim 0.1$; $ \alpha_{\rm cold} \sim 0.01$) to
successfully reproduce dwarf nova outbursts.

For a disk to be in a globally hot-stable regime, the accretion rate
must be everywhere larger than the value of $\dot{M}_{\rm B}$ at the
outer edge. Similarly, to be in a globally cold-stable state, the
accretion rate must be everywhere smaller than the value of
$\dot{M}_{\rm A}$ at the disk inner radius. If neither of these
criteria is satisfied, then parts of the disk are partially ionized
and subject to instability, which results in large amplitude
outbursts.

The X-ray luminosities of candidate CVs in GCs are useful in that they
provide constraints on the rate at which mass is being accreted onto
the white dwarf in these systems. Boundary layers in CVs accreting at
low rates ($\lta 10^{16}$~g~s$^{-1}$) should be hot, optically-thin
and emitting X-rays {\citep{tylen,narpop,medmen,psav}}. Using the
standard virial argument for half of the energy being released in the
boundary layer, the X-ray luminosity is roughly expected to be
\begin{equation}
L_X = 1.33 \times 10^{32}\eta_X {\dot M_{\rm in} \over 10^{15} \rm
g\ s^{-1}} {M_1 \over \rm M_\odot} \left({r_{\rm in} \over 10^{9}
\rm cm} \right)^{-1} \rm erg\ s^{-1}, \label{lx}
\end{equation}
where $\eta_X \leq 1$ is the efficiency of conversion of the energy
released into X-rays and $\dot M_{\rm in}$ is the accretion rate at
the disk inner radius, $r_{\rm in}$ (assumed to extend all the way
down to the white dwarf, for now). Given the uncertain efficiency
$\eta_X$, observed X-ray luminosities provide only approximate lower
limits on $\dot M_{\rm in}$.  Still, it follows rather simply from the
X-ray-inferred accretion rates and the DIM stability criteria that
candidate CVs in GCs should typically be subject to dwarf nova
outbursts if they contain extended disks reaching down to the white
dwarf's surface (see \S\ref{results} for detailed criteria). This
would also imply that, at any time, most of these candidate CVs should
be in a quiescent phase of accretion, during which mass is being
accumulated in the disk, until the next outburst is triggered.  By
comparison with the outburst properties of CVs in the field, however,
one is forced to recognize that candidate CVs in GCs do indeed appear
to be erupting very infrequently.

\section{A metallicity effect?}

One of the well-known characteristics of stellar systems in GCs is
their lower average metallicity {compared to} systems in the field:
could this be the underlying cause of rare cataclysms in GCs?
Following the conclusions of \citet[][see also Menou 2000]{gammen}, it
seems possible that dwarf nova outbursts are effectively triggered
when a critical ionization rate is reached that is sufficient for MHD
recoupling, turbulence and transport to develop fully in the disk, in
conditions approaching ideal MHD. This is likely to occur at much
lower values of the ionization fraction (perhaps $\sim 10^{-5}$;
Gammie \& Menou 1998) than the fraction $\sim 0.5$ at which ionization
affects opacities: this low critical ionization value for MHD
turbulence appears to be the relevant ``master'' trigger for the
outbursts, which would then be followed by disk heating and then
opacity changes (as postulated in the DIM) when the ionization
fraction approaches $\sim 0.5$.

In the bulk of the quiescent disk, thermal first--ionization of alkali
metal is the dominant mechanism contributing to the low level of
ionization required \citep{gammen}. Given the exponential dependence
on temperature of the ionization fraction in the Saha equation,
relative to the linear dependence on electron density, it appears very
unlikely that a change by even one or two orders of magnitude in the
metallicity of the gas can affect much the condition for outburst
triggers, which will be entirely dominated by what sets the
temperature conditions in the disk. This indicates that, in the
unsteady evolving disk postulated to be present during quiescence in
the DIM, metallicity is not expected to have much of an effect on the
conditions triggering outbursts.

If the structure of quiescent dwarf nova disks were to be layered in
quiescence, however, the situation could be quite different
\citep{gam96,men01}. In the simplest version of the layered accretion
scenario, the temperature of the isothermal dead zone is constant and
determined by the level of activity of the MHD-turbulent surface
layers. The only time-dependent component in the model is the gradual
accumulation of mass in the dead zone. At a fixed temperature but
increasing density, one can imagine how the metallicity of the gas
being accreted in a given system can be important in determining the
exact level of ionization present in the disk and thus when the next
outburst will be triggered. As it turns out, however, mass
accumulation alone cannot trigger an outburst through MHD
recoupling. Using for simplicity the Saha equation for a
first-ionized, single species gas (e.g. the most easily ionized alkali
metal), and fixing the temperature, we find that in the low ionization
limit ($n_e/n_n \ll1$), the ionization fraction $n_e/n_n$ varies with
density as $1/\sqrt{n_n}$, where $n_e$ and $n_n$ are the electron and
neutral densities, respectively.  The same scaling obtains for the
magnetic Reynolds number, which goes down as a result of mass
accumulation in the dead zone. The importance of Hall terms also goes
down relative to the Ohmic terms at larger densities
\citep[][e.g.]{balter}. Therefore, even though the absolute value of
the ionization fraction would depend linearly on metallicity in this
simple fixed-temperature model for layered accretion, the ionization
fraction would go down with time as mass accumulates in the dead zone
and there would be no way of triggering an outburst on the basis of
MHD recoupling at later times.

We also note that if one were to invoke the possibility of finite
stresses and dissipation in the dead zone \citep{flesto}, this layer
would then no longer be isothermal and at a fixed temperature as gas
accumulates. The exponential dependence on temperature of the
ionization fraction is then again very likely to overcome any linear
metallicity dependence. We conclude from these various considerations
that, according to existing models for quiescent dwarf nova disks, it
does not appear that the low metallicity of donors in GCs is the
underlying cause of their rare cataclysms.

One remaining possibility related to low metallicities is that the
quality of magnetic coupling in the donor's atmosphere and thus the
amount of magnetic flux being transferred to the disk are actually
able to influence the disk evolution by modifying its MHD conditions,
as proposed by \citet{MMH99}. If it were the case that, on average,
the material being transferred in GCs carries less magnetic flux than
in field CVs, it could in principle explain the reduced activity of
candidate CVs in GCs. However, since the influence of metallicity on
the magnetic properties of a stellar atmosphere or the resulting
effect that this would have on the disk MHD behavior are not
understood at a quantitative level, it is currently difficult to rule
in favor or against this scenario.

\section{Low mass transfer rates?} \label{sec:low}

In this section, we want to determine whether it is possible for
candidate CVs in GCs to be quiescent dwarf novae in which the donor
star transfers mass at such a low rate that outbursts are much rarer
than in field CVs. Although the direction of this effect is obvious,
its magnitude is not. We have therefore decided to quantify how the
outburst properties of a representative system depend on the mass
transfer rate with a series of detailed time-dependent disk
instability models. We have not explored scenarios in which
differences between field and GC CVs are attributed to different
values of the viscosity parameter $\alpha$. If there is a
``universal'' saturation value for the turbulent transport driving
accretion in these disks, it would presumably not differ between field
and GC CVs.

It should be noted that, to the best of our knowledge, there has not
been any systematic study of the observational statistics of dwarf
nova outbursts in GCs.  The quantity of most interest probably is the
duty cycle of the outbursts. In principle, given a number of candidate
CVs, a typical duty cycle, and the statistics of observations for a
given cluster, one can predict the number of dwarf nova outbursts one
ought to be observing at any time or through repeated observations
(modulo the small number statistics). Additional outburst diagnostics
may also be useful in determining the regime of accretion present in
candidate CVs in GCs. We hope that the detailed study of outburst
properties presented here will motivate a statistical study of
observed duty cycles and perhaps new observational programs to obtain
additional constraints.

\subsection{The numerical model}

In this section we use a standard version of the DIM, in which the
disk extends all the way down to the white dwarf's surface, the mass
transfer rate from the secondary star is constant and effects of disk
irradiation are neglected. The numerical model, described in detail in
\citet{Hametal98}, follows the thermal-viscous evolution of a
geometrically thin disk around the white dwarf. The equations of
conservation of mass, angular momentum and energy for a Keplerian disk
are solved on an adaptive grid which resolves narrow structures in the
disk. A grid of disk vertical structures, which determine the local
cooling rate of the disk as a function of its surface density, central
temperature and vertical gravity, is precalculated before running the
disk evolution. We only use models after they have converged to a
periodic outburst pattern: the results shown are robust in the sense
that memory of the initial conditions has been lost.

We show results for a specific model with parameters close to
appropriate for the field dwarf nova VW Hyi ($P_{\rm orb} \simeq
1.8$~hr). The model has the following parameters: primary mass
$M_1=0.6 \rm M_{\odot}$, time-averaged value of the disk outer radius
$<R_{\rm out}>= 2 \times 10^{10}$ cm, value of the disk inner radius
$R_{\rm in}= 8.5 \times 10^{8}$ cm and values of the viscosity
parameters (unless otherwise specified) $\alpha_{\rm hot} =0.2$ and
$\alpha_{\rm cold} = 0.04$.  The numerical resolution adopted is N=800
radial grid points.

\subsection{Varying the disk inner radius $R_{\rm in}$}

In the DIM, the value of the disk inner radius ($R_{\rm in}$, which is
also the white dwarf stellar radius) does have some influence on the
outburst properties of an unstable disk. Therefore, before launching
into an extensive study of the dependence on the mass transfer rate,
we have conducted a series of tests to determine the magnitude of the
$R_{\rm in}$ dependence.

\begin{figure}
\plotone{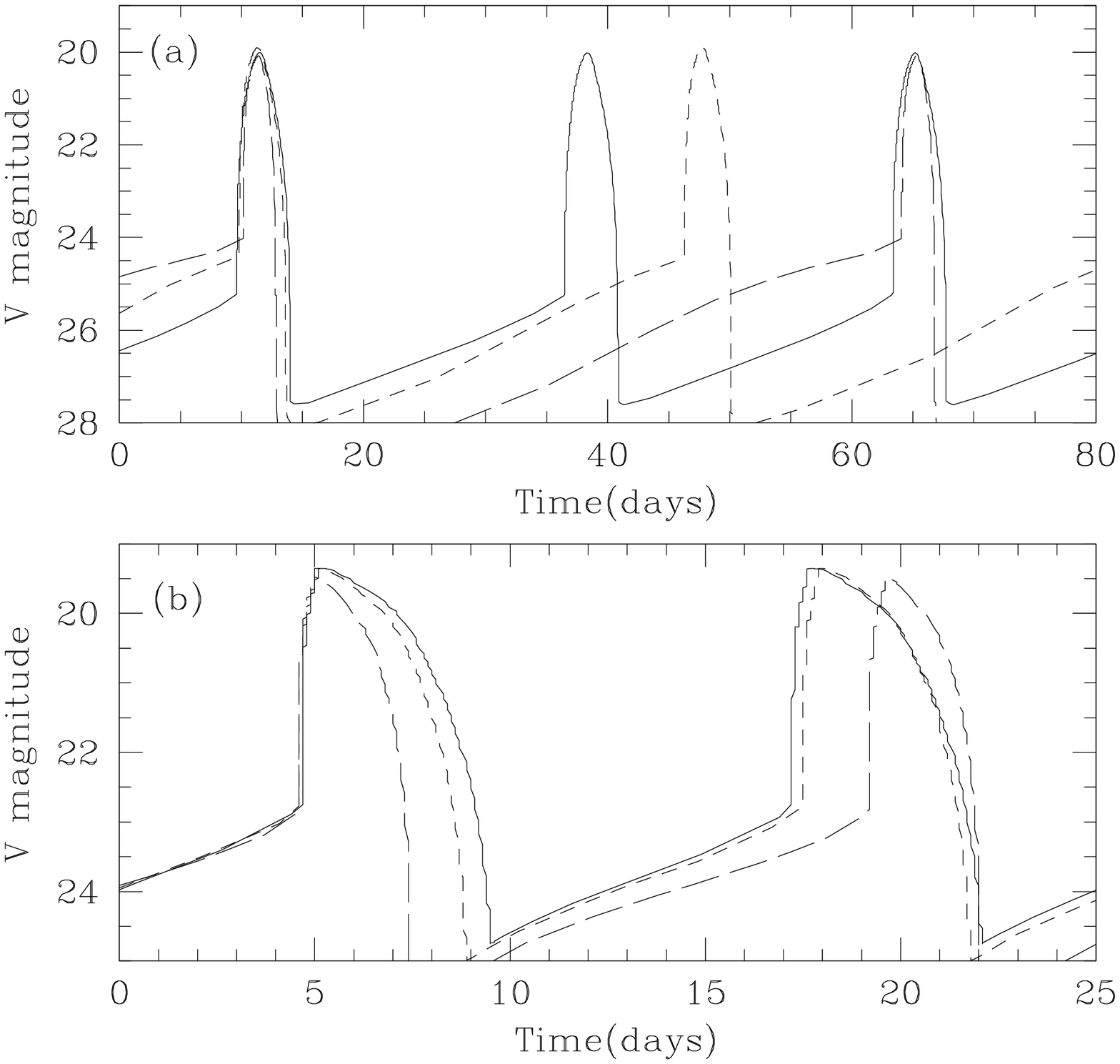}
\caption{Outburst cycles predicted in a model with $\dot M_T = 1
\times 10^{16}$ g s$^{-1}$ (a) and $7.5 \times 10^{16}$ g s$^{-1}$
(b). In each panel, lightcurves for the following values of the disk
inner radius are shown: $R_{\rm in}= 8.5 \times 10^8$~cm (solid line),
$R_{\rm in}= 11.5 \times 10^8$ cm (short-dashed line), $R_{\rm in}= 28
\times 10^8$ cm (long-dashed line). For clarity, lightcurves have been
aligned on the first outburst peak.}
\label{fig:one}
\end{figure}

Figure~\ref{fig:one} shows V-band lightcurve predictions from models
with different $R_{\rm in}$ values, at a given mass transfer rate,
$\dot M_T$.  Figure~\ref{fig:one}a shows models with $\dot M_T=1
\times 10^{16}$ g s$^{-1}$, for which outbursts are triggered in the
innermost regions of the disk (``inside-out outbursts'').
Figure~\ref{fig:one}b shows models with $\dot M_T=7.5 \times
10^{16}$ g s$^{-1}$, for which outbursts are triggered in the
outermost regions of the disk (``outside-in outbursts''). In each
panel, the solid line corresponds to a model with $R_{\rm in}= 8.5
\times 10^8$ cm, the short-dashed line to $R_{\rm in}= 11.5 \times
10^8$ cm and the long-dashed line to $R_{\rm in}= 28 \times 10^8$ cm.

The recurrence times predicted in the model shown in panel (a) depend
on the value of $R_{\rm in}$ because inside-out outbursts are
triggered close to the disk inner edge, and the value of the critical
surface density, $\Sigma_{\rm max}$ \citep[$\propto R^{1.1}$; see,
e.g.,][]{Hametal98}, at which mass accumulation triggers the
thermo-viscous instability, increases with increasing $R_{\rm in}$
(the ignition radius $R_{\rm ignit}$ is $\simeq 1.5 \times R_{\rm in}$
in each one of these models). On the contrary, the recurrence time is
basically independent of $R_{\rm in}$ in the model with outbursts
triggered in the outer regions of the disk (panel (b); $R_{\rm ignit}
\simeq 9 \times 10^{9}$ cm in all 3 cases).

Models in Fig.~\ref{fig:one}a, with a mass transfer rate on the low
side and inside-out outbursts, are probably the most relevant ones for
candidate CVs in GCs.  Even when multiplying the disk inner radius by
a factor of three, however, the effect is only a doubling of the
outburst periodicity. Despite uncertainties associated with unknown
white dwarf masses in GCs (which may be systematically larger than in
the field because of mass segregation), the corresponding
uncertainties for disk inner radii will remain small compared to the
large factor of three variation shown in
Fig.~\ref{fig:one}a. Therefore, we conclude from these tests that
fixing the disk inner radius to a reasonable value, $R_{\rm in}= 8.5
\times 10^8$~cm, is unlikely to affect much our conclusions on the
effects of variations in the mass transfer rate.

\subsection{Varying the mass transfer rate $\dot M_T$}

We have sampled a range of mass transfer rates from $\dot M_T
=10^{15}$ g s$^{-1}$ to $\dot M_T \simeq 2 \times 10^{17}$ g s$^{-1}$
with about ten models. Above a few $10^{17}$ g s$^{-1}$, the disk
becomes globally stable. For each model, three different $\alpha_{\rm
hot}$/$\alpha_{\rm cold}$ viscosity prescriptions were
investigated. Each model produces a lightcurve of the type shown in
Fig.~\ref{fig:one}. The morphology of these lightcurves changes with
$\dot M_T$ and the $\alpha$ values adopted. In order to quantify and
better characterize the lightcurve properties, we have defined nine
useful attributes that we have systematically measured for all the
calculated model lightcurves. As we elaborate further below, some of
these attributes are direct observables, and as such, they could allow
further observational tests of a given regime of accretion in CVs.

\subsubsection{Characterizing outburst cycles}

Peak V mag. -- This is the V-band magnitude of the disk when it is
most luminous (at the peak of an outburst). The V-band magnitude has
been calculated by integrating the multi-temperature blackbody
emission from the disk (magnitude scale arbitrary).

$R_{\rm ignit}$ -- This is the ignition radius, i.e. the radius of the
disk annulus which first becomes ionized, unstable and subsequently
triggers a global outburst in the disk.

$T_{\rm cycle}$ -- The recurrence (or cycle) time is computed as the
time between two successive lightcurve peaks.

$T_{\rm duration}$ -- The outburst duration time is computed as the
time during which the disk V-band magnitude is within 3 magnitudes of
its peak value. We have checked that this definition ensures that
$T_{\rm duration}$ is estimated at a value very similar to what an
eyeball estimate would give. Also, with this definition, we do not
have to worry about defining properly a quiescence level, which would
likely depend on contributions from a hot spot and the secondary
star. $T_{\rm duration}$ is the sum of $T_{\rm rise} + T_{\rm decay}$
(defined below).

Duty cycle -- It is simply calculated here as the ratio $T_{\rm
duration}$/$T_{\rm cycle}$. It basically captures the fraction of the
time the dwarf nova is ``on'' (given our specific definition of
$T_{\rm duration}$).

$T_{\rm quiescence}$ -- It is calculated as the difference $T_{\rm
cycle} - T_{\rm duration}$, where $T_{\rm duration}$ is taken for the
outburst following the corresponding quiescence phase.

$T_{\rm rise}$ -- The rise time is calculated as the time taken by the
disk V-band magnitude to rise from its peak magnitude + 3 to that peak
magnitude.

$T_{\rm decay}$ -- The decay time is calculated as the time taken by
the disk V-band magnitude to decrease from its peak magnitude to that
value + 3. Note that this definition includes any ``plateau'' phase in
the decay, which are found to exist at large values of the mass
transfer rate, $\dot M_T$.

V fluence -- This is the fluence {(time-integrated luminosity) of
an outburst in the V band, calculated over the estimated outburst
duration time $T_{\rm duration}$ (i.e. covering a range of 3
magnitudes around the peak V value).} Here again, the absolute energy
scale is arbitrary.

Note that, as is evident from their definitions, all these quantities
are not independent from each other.

\subsubsection{Results}

\begin{figure*}
\plotone{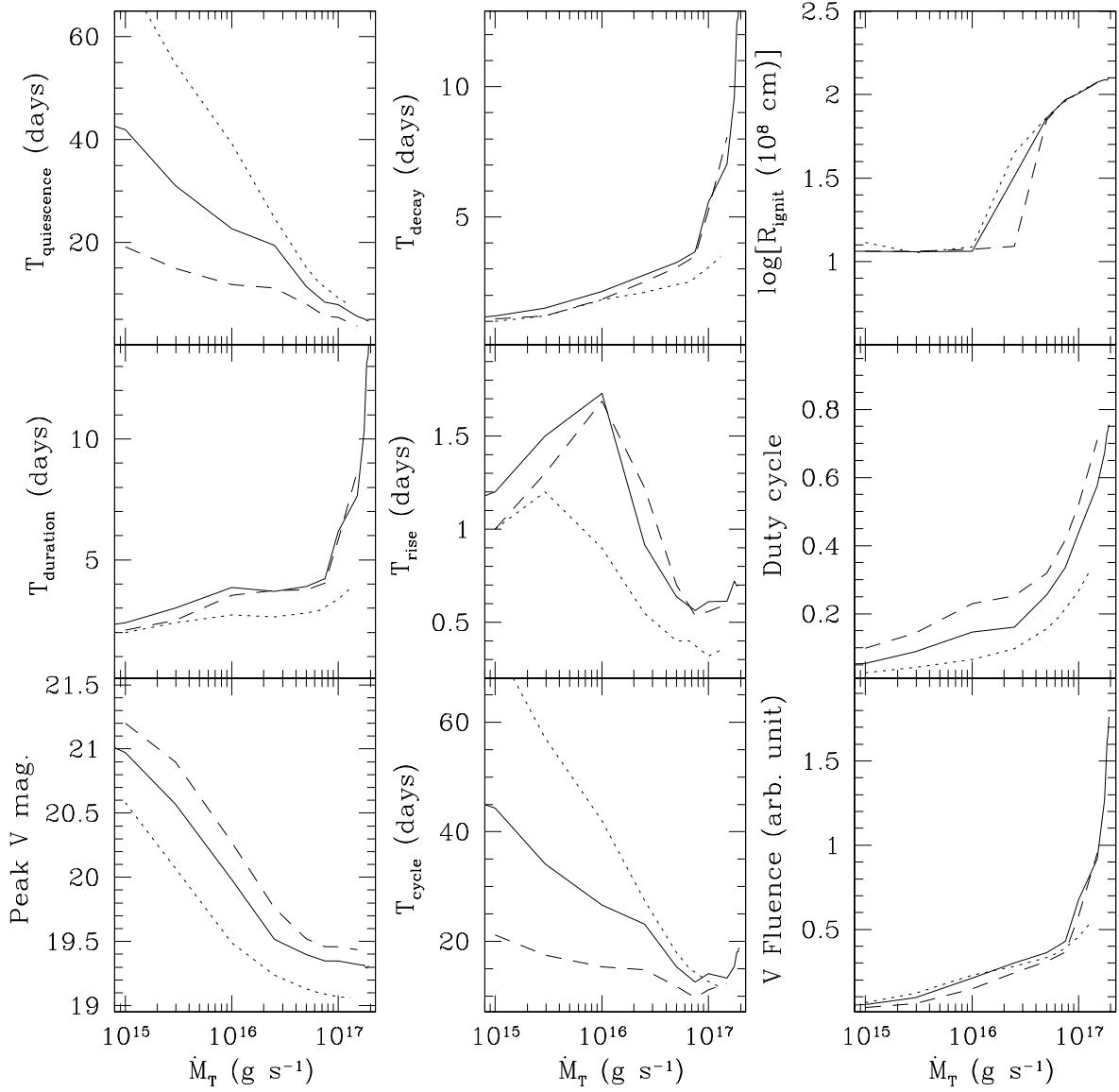}
\caption{Predicted variations in outburst cycle properties as a
function of the mass transfer rate, $\dot{M}_T$, from $ 10^{15}$ to $2
\times 10^{17}$ g s$^{-1}$. In each panel, the solid line corresponds
to the viscosity prescription ($\alpha_{\rm hot}$, $\alpha_{\rm
cold}$)=(0.2, 0.04), the dotted line to ($\alpha_{\rm hot}$,
$\alpha_{\rm cold}$)= (0.3, 0.04) and the dashed line to ($\alpha_{\rm
hot}$, $\alpha_{\rm cold}$)= (0.2, 0.05). See text for details.}
\label{fig:two}
\end{figure*}

Figure~\ref{fig:two} shows how these nine outburst cycle attributes
vary as a function of the mass transfer rate, $\dot M_T$. Solid, dashed
and dotted lines in each panel show results for three different
viscosity prescriptions: ($\alpha_{\rm hot}$, $\alpha_{\rm
cold}$)=(0.2, 0.04), (0.2, 0.05) and (0.3, 0.04), respectively.
Changes in outburst attributes from one to the other of these
prescriptions is only quantitative in nature. The qualitative trends
with $\dot M_T$ shown in Fig.~\ref{fig:two} are therefore robust with
respect to uncertain $\alpha$ prescriptions.

Note that we have not explored dependencies in our models related to
the value of the disk outer radius, $R_{\rm out}$, which is itself
determined by the binary's orbital period, $P_{\rm
orb}$. \citet{smak99} has argued in a related analysis that the
dependence of the duration time, $T_{\rm duration}$, on the orbital
period should scale as $P_{\rm orb}^{0.74}$. According to this
scaling, even for a change by a factor 3 in orbital period, the
corresponding change in $T_{\rm duration}$ (and possibly other
outburst time attributes, such as the duty cycle) would only be about
a factor of 2. This is comparable to uncertainties related to our poor
knowledge of $\alpha$, according to the results shown in
Fig.~\ref{fig:two}. For this reason, we have chosen to focus our
numerical exploration on variations of $\dot M_T$ at a fixed disk size
(i.e. same time-averaged value of $R_{\rm out}$). In the future, it
may be interesting to explore further this dependence of model
predictions on the binary's orbital period, especially if there were
reasons to believe that candidate CVs in GCs have a different period
distribution than CVs in the field.

As $\dot M_T$ is decreased, we find that the outburst fluence, its
duration time $T_{\rm duration}$, its decay time $T_{\rm decay}$ and
the duty cycle all decrease, while the quiescence time $T_{\rm
quiescence}$, the recurrence time $T_{\rm cycle}$ and the outburst
peak magnitude increase. The variations in the ignition radius,
$R_{\rm ignit}$, and the rise time, $T_{\rm rise}$, both reflect the
transition from outside-in to inside-out outbursts occurring at a mass
transfer rate of a few $10^{16}$ g s$^{-1}$ (see also
Fig.~\ref{fig:one}). This transition is characterized by a significant
increase in the rise time of the outburst (in the V band). At the
highest accretion rates, the existence of extended plateau phases
during outburst becomes obvious as the values of, e.g., $T_{\rm
decay}$ and the V-band fluence, become large.

The trends with $\dot M_T$ shown in Fig.~\ref{fig:two} are strongly
correlated with each other, and as such, they could be used to
statistically test the notion that mass transfer rates are, on
average, smaller in the population of CVs found in GCs, than in the
field. As shown by Fig.~\ref{fig:two}, this would require shorter
duration and decay timescales for the outbursts, less luminous
outbursts and reduced fluences (in the V band). Although it would be
dangerous (and probably misleading) to do a comparison between
individual systems, we believe that the trends shown in
Fig.~\ref{fig:two} are rather robust and therefore meaningful in a
global statistical sense. One would expect them to be followed by the
few CVs showing dwarf nova outbursts in GCs, if it is true that their
disks are fed, on average, at much lower rates than CVs in the field.

Perhaps the most easily used attribute which is shown in
Fig.~\ref{fig:two} is the duty cycle of the outbursts. At mass
transfer rates $\dot M_T \lta 3 \times 10^{15}$~g~s$^{-1}$, a duty
cycle $\sim 0.1-0.025$ obtains, depending on the $\alpha$ values
adopted. It is already interesting to ask whether these values are
consistent with the paucity of dwarf nova outbursts in a GC such as 47
Tuc. With 22 confirmed CVs and this type of values for the duty cycle,
one would expect about 2-0.5 dwarf novae to be active (i.e. in
outburst) at any time when 47 Tuc is observed!  To do this comparison
properly requires taking into account effects such as source crowding,
the weaker nature of the outburst expected (see values of peak V
magnitudes in Fig.~\ref{fig:two}), etc...but this simple scaling
already suggests that mass transfer rates $\dot M_T \lta 3 \times
10^{15}$~g~s$^{-1}$ may not necessarily easily explain the paucity of
outbursts in at least one well-studied GC. The possibility that
several tens of additional X-ray sources in 47 Tuc are in fact
unconfirmed CVs \citep{towbil,Heinke05} would make this comparison
even more meaningful.

In addition, it is important to note that one is not free to pick low
values of $\dot M_T$ to justify a low frequency of outbursts in GCs
and posit at the same time that the observed low luminosity X-ray
sources are {the} same systems, seen as dwarf novae in a quiescent
phase. Indeed, the DIM requires the accretion rate in a quiescent disk
not to be constant but rather fall with decreasing radius at least as
steeply as the critical scalings given in
Eqs~(\ref{mdmax1}--\ref{mdmin1}). For the specific model we are
considering here, if the disk is fed at a rate $\dot M_T =
10^{15}$~g~s$^{-1}$, the implied accretion rate onto the white dwarf
is $\dot M_{\rm in} \lta 10^{12}$~g~s$^{-1}$, i.e.  so low that it
becomes essentially impossible to power the observed X-ray
luminosities even with a $100 \%$ efficient X-ray conversion (see
Eq. \ref{lx}). This limitation is the main reason why we did not
extend our detailed models below values of $\dot M_T =
10^{15}$~g~s$^{-1}$ in Fig.~\ref{fig:two}.

By extension, given a distribution of X-ray luminosities for a
population of candidate CVs in a GC such as 47 Tuc, one would have to
pick a distribution of mass transfer rates which is compatible with
these X-ray luminosities. Combining this constraint with the allowed
values of the duty cycle shown in Fig.~\ref{fig:two} suggests that it
would be rather difficult to explain the paucity of dwarf nova
outbursts in GCs by invoking only very low mass transfer rates and
stay consistent with the observed X-ray luminosities. Clearly, it
would be interesting to derive in the future an improved, more
quantitative version of this argument by using a better sampled
statistics on the frequency (and perhaps additional attributes) of
dwarf nova outbursts in a given GC.

\subsection{Very long recurrence times ?}

We have thus argued that it would be difficult for the population of
candidate CVs in GCs, if they possess fully extended disks, to
experience outbursts with recurrence times that would be much longer
than those in the population of CVs in the field.  We reached this
conclusion using the DIM with standard parameters, in particular
viscosity parameters $\alpha_{\rm hot} \gta 0.2$ and $\alpha_{\rm
cold} \gta 0.04$. Although the value of $\alpha_{\rm hot}$ appears to
be universal \citep{smak99}, it may not be the case for $\alpha_{\rm
cold}$. Indeed, there exist dwarf novae in the field with very long
recurrence times, the best known example of which is WZ Sge (with a
recurrence time $\sim 30$ years). This class of systems has been
mentioned as a possible explanation for rare outbursts in GCs
\citep{G99}.  {For these} exceptionally long recurrence times (and the
long outburst durations), {the DIM requires} a very low value of
$\alpha_{\rm cold}\sim 3 \times 10^{-5}$ \citep[][]{smak93,O95}. Such
very low values of the efficiency of angular momentum transport in
quiescent disks would require a physical explanation. \citet{MMH99}
have suggested that the viscosity in quiescent accretion disks is
determined by the amount of magnetic flux transferred from the
companion star together with the matter being accreted. According to
this suggestion, when the companion leaves the main sequence to become
degenerate, near the CV period minimum, its atmospheric temperature
becomes so low that the amount of magnetic flux transferred is much
reduced, resulting in much lower $\alpha_{\rm cold}$ values in
quiescent disks. Since WZ Sge is very close to the CV minimum period,
this could explain its extremely low viscosity in quiescence. To
extend this interpretation to candidate CVs in GCs, however, would
require most of them to have ``brown dwarf" companions. At this point,
observed CV periods in GCs do not support this expectation \citep[see
e.g.][]{Ed03b}.

An alternative solution to the WZ Sge long recurrence time puzzle,
which does not involve any unusually low $\alpha_{\rm cold}$ value,
has been proposed by \citet{Letal95} \citep[see
also][]{Hametal97,Letal99}. Noting that inner truncation can stabilize
an otherwise unstable disk, these authors suggested that the quiescent
disk in WZ Sge may be truncated by the magnetic field from the white
dwarf or by ``evaporation'' of the innermost disk regions \citep[see
e.g][]{MMH94}, sufficiently so that the disk becomes only marginally
unstable. As a result of the truncation, one also expects an accretion
rate at the inner edge of the (truncated) disk which is consistent
with the observed X-ray luminosity.  This model requires a rather
strong enhancement of the mass transfer rate during outburst to
reproduce observed outburst properties, however. This point is subject
to theoretical controversy \citep[][and reference therein]{smak04} but
it seems to be supported by observations \citep{ste04}.  Whether this
explanation for the very long recurrence times of WZ Sge is correct or
not, the idea of disk stabilization by truncation is worth exploring
for candidate CVs in GCs, as it seems to be a reasonable explanation
for their resounding silence.

\section{Stabilized disks in globular clusters?} \label{results}

Most CVs in GCs could contain stable disks, which are not subject to
the thermo-viscous instability responsible for outbursts in field CVs,
if regions of their disks which would normally be subject to the
instability are systematically truncated. In this section, we first
show that it is not possible to have such stable disks in candidate
CVs in GCs unless truncation of the disk inner regions indeed happens.
Then, we quantify the conditions under which the magnetic field from
the accreting white dwarf is sufficiently strong to achieve
stabilization of the disk by magnetospheric truncation. Given the
unknown orbital period, nature of the donor star and composition of
the material being transferred and accreted in these systems, we
consider a number of possibilities successively below. In each case,
we directly compare the values of the mass transfer rate, $\dot M_{\rm
tr}$, required to satisfy the DIM global stability criteria,
Eq.~(\ref{mdmax1}-\ref{mdmin1}), to the typical values of $\dot M_{\rm
tr}$ necessary to power representative X-ray luminosities $L_{\rm X}
\sim 10^{31}$ and $10^{32.5}$~erg~s$^{-1}$ according to
Eq.~(\ref{lx}), for $\eta_X=0.5$. We adopt the white dwarf mass-radius
relation ($M_{\rm WD}$--$R_{\rm WD}$) of \cite{nau} for the accretor
whenever it is necessary for our calculations.

\subsection{Full disk, main sequence secondary}

\begin{figure}
\plotone{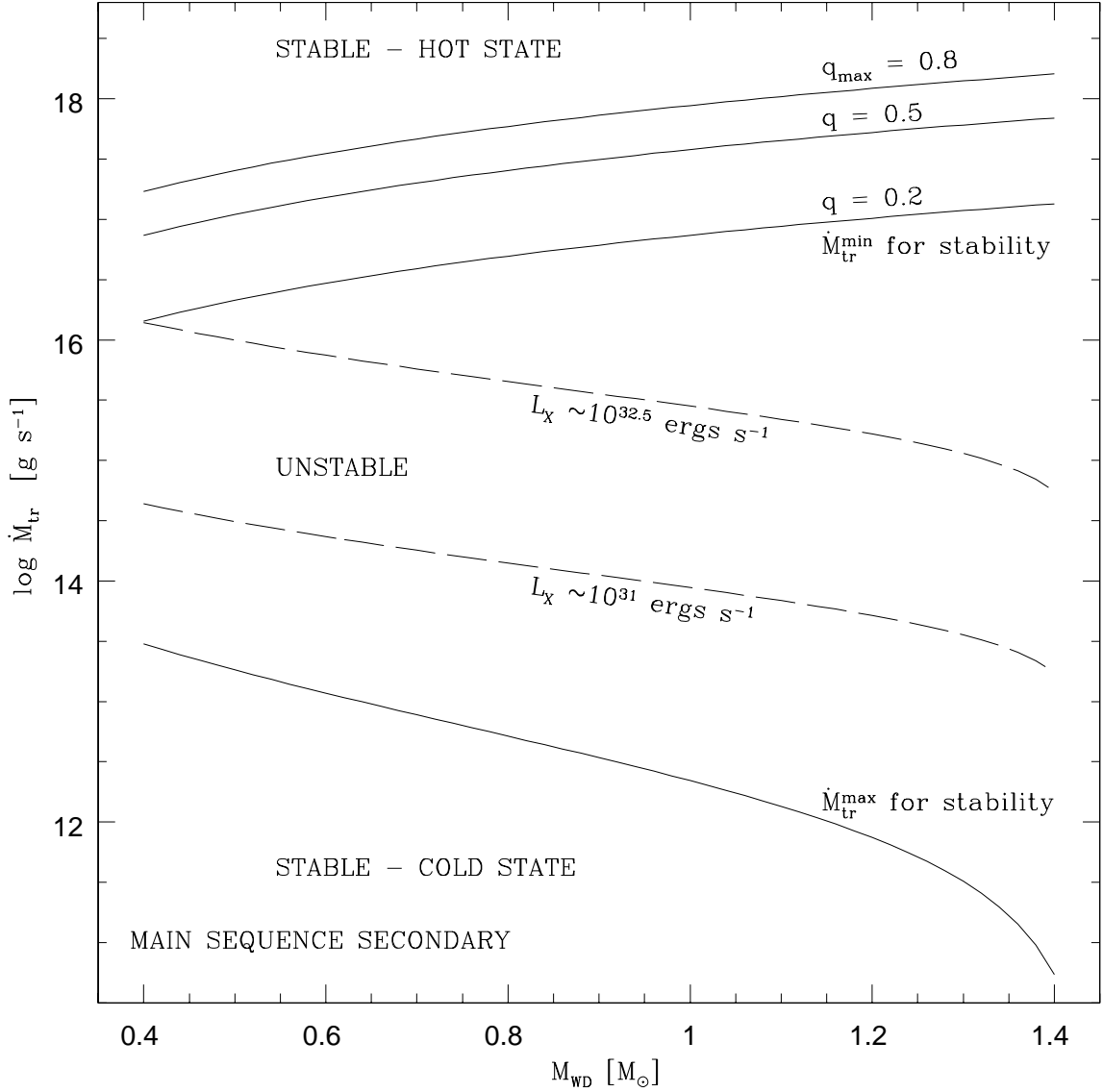}
\caption{ Comparison of the values of the mass transfer rate, $\dot
M_{\rm tr}$, required for a full CV disk to be globally hot and stable
(above the upper solid lines) or globally cold and stable (below the
lower solid line), with the values necessary to power X-ray
luminosities $L_{\rm X} \sim 10^{31}$ and $\sim
10^{32.5}$~erg~s$^{-1}$ (long-dashed lines) according to
Eq.~(\ref{lx}), for $\eta_X=0.5$. White dwarf masses in the range
$M_{\rm WD}=0.4$-$1.4 M_\odot$ are considered. The donor is assumed to
be a main-sequence star.  }
\label{fig:three}
\end{figure}

As mentioned in \S\ref{stable}, for the disk to be in a globally hot
and stable state, the mass-transfer rate must exceed the value of the
accretion rate $\dot M_{\rm B}$ given by Eq.~(\ref{mdmin1}) at the disk
outer radius. We adopt for simplicity a value of the disk outer radius
$R_{\rm D}= 0.9~R_{L1}$, where $R_{L1}$ is the primary's mean Roche-lobe
radius,
\begin{equation}
R_{L1} = 0.462~a \left( \frac{M_{\rm 1}}{M_{\rm 1} + M_{\rm 2}}
\right)^{1/3},
\label{roche}
\end{equation}
where $M_{\rm 2}$ is the mass of the secondary and $a$ is the binary
orbital separation, determined by Kepler's law
\begin{equation}
a = 3.5 \times 10^{10} \left( \frac{M_{\rm 1}}{M_{\rm \odot}}
\right)^{1/3} (1+q)^{1/3}~P_{\rm hr}^{2/3}~{\rm cm},
\end{equation}
with $q = M_{\rm 2}/M_{\rm 1}$ and $P_{\rm hr}$ is the orbital period
in hours. For the main-sequence mass-period relation, we use $m_{\rm
2} \simeq 0.11~P_{\rm hr}\simeq R_2/R_{\odot}$ \citep[see
e.g.][]{king88}, where $m_2=M_2/M_{\odot}$. The condition
$\dot{M}_{\rm tr}= \dot M_{\rm B} (R_{\rm D})$ for values of the
mass-ratio $q=0.2$, $0.5$ and $0.8$ are shown as a function of the
primary's mass, $M_{\rm WD}$, as the three upper solid lines in
Fig.~\ref{fig:three}. Above these lines, the disk is hot and stable.

For a disk to be in a globally cold and stable state, the mass
transfer rate cannot exceed the value of the accretion rate given by
$\dot M_{\rm A}$ in Eq.~(\ref{mdmax1}) at the disk inner edge, which
is also the white dwarf radius in the case of a fully extended disk.
The condition $\dot{M}_{\rm tr}= \dot M_{\rm A} (R_{\rm WD})$ is shown
as a function of the primary's mass, $M_{\rm WD}$, as the lower solid
line in Fig.~\ref{fig:three}. Below this line, the disk is cold and
stable.  In addition, the two long-dashed lines in
Fig.~\ref{fig:three} show, as a function of the primary's mass, the
typical mass transfer rates required to power X-ray luminosities
$L_{\rm X} \sim 10^{31}$ and $10^{32.5}$~erg~s$^{-1}$ according to
Eq.~(\ref{lx}) for $\eta_X=0.5$.

{As already emphasized by \citet{Sharetal96} and \citet{Ed03b},
globally hot and stable disks would be too bright to explain the faint
optical luminosities of candidate CVs in GCs. These candidate CVs also
appear} much too X-ray bright to contain globally cold and stable
disks extending all the way to the white dwarf's surface. From the
comparison shown in Fig.~\ref{fig:three}, one would naively expect
these candidate CVs in GCs to contain unstable disks and thus show
dwarf nova outbursts much like their counterparts in the field.

\subsection{Full disk, degenerate hydrogen-rich secondaries} \label{hdeg}

\begin{figure}
\plotone{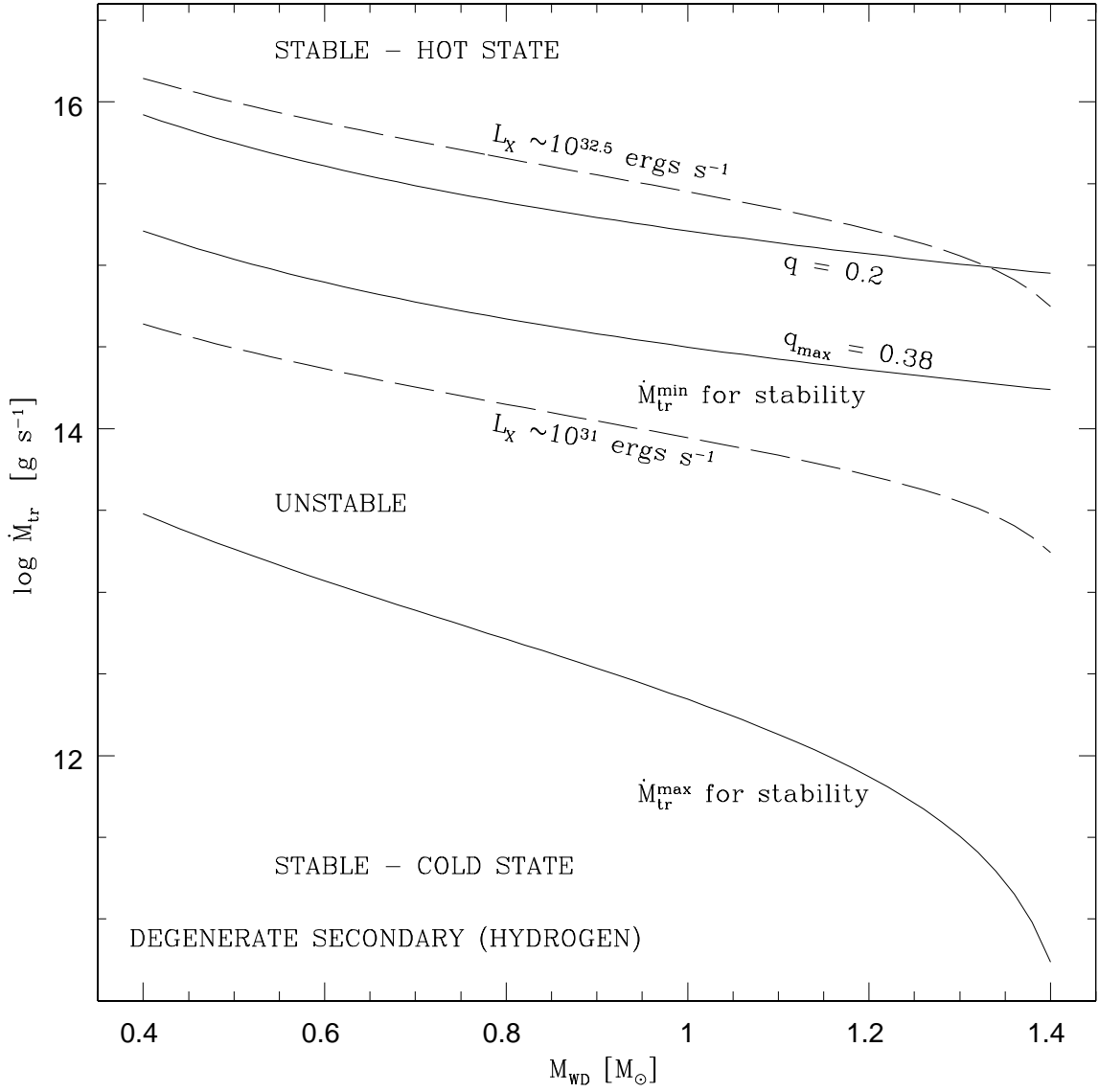}
\caption{Same as Fig.~\ref{fig:three} when the donor is assumed to be
a hydrogen-rich degenerate star.}
\label{fig:four}
\end{figure}

Field CVs are supposed to evolve through a period minimum at $\lta
80$~min, at which point they become degenerate \citep[see
e.g.][]{king88}.  Mass-transfer rate for such post-minimum systems are
very low and if such systems were to exist in globular clusters, they
would be interesting candidates for CVs containing cold and globally
stable accretion disks. To test this hypothesis, we modify our
analysis and assume the following mass-radius relation for the
secondary \citep{king88}
\begin{equation}
R_{\rm 2} = 10^9 \left(1+X \right)^{5/3} m_2^{-1/3} \rm cm,
\label{deg1}
\end{equation}
which leads to a mass-period relation \beq m_2= 1.5 \times 10^{-2}
\left(1+X \right)^{5/2}P_{\rm h}^{-1}.  \eeq The results are shown in
Fig.~\ref{fig:four}, where $X=0.7$ was assumed.  Although some of the
candidate CVs in GCs with X-ray luminosities at the high end of the
observed luminosity function could be described as systems containing
hydrogen-rich degenerate donors and globally hot and stable disks,
this could not be a solution for the entire population {unless one is
willing to consider extremely low values of the X-ray conversion
efficiencies, $\eta_X$. In addition, the faint optical luminosities of
candidate CVs in GCs appear to be inconsistent with globally hot and
stable disks \citep{Sharetal96,Ed03b}.}

\subsection{Full disk, degenerate helium secondaries} \label{hedeg}

One should also consider the possibility that the donors in candidate
CVs in GCs are helium-rich WDs, i.e. that these systems are members of
the AM CVn class of double degenerate binaries \citep{Warner95,Nelm}.

For our purpose, we will simply assume the mass-radius relation
Eq.~(\ref{deg1}) with X=0 and rederive the corresponding stability
criteria.  More realistically, the mass-radius relation for donors in
AM CVn binaries would depend on their formation channel
\citep{deloye}, which could be rather complex in globular clusters
\citep[see e.g.][]{ivanova}.

One also has to modify the global stability criteria for disks which
are now composed primarily of helium, with ionization properties which
differ substantially from those of hydrogen. The criterion for a disk
to be hot and stable becomes \citep{menperna}
\begin{equation}
\dot{M}_{\rm B} = 5.9 \times 10^{16} \left( \frac{\alpha}{0.1}
\right)^{0.41} m_1^{-0.87} \left( \frac{r}{10^{10}\ \rm cm}
\right)^{2.62} ~\rm g~s^{-1}.
\end{equation}
The value of $\dot{M}_{\rm A}$ for a disk to be cold and stable was
derived following the same algorithm as \citet{smak83} or
\citet{tsug}. The stability condition in this case is obtained by
requiring that the maximal temperature in a cold and stable disk
($R^T_{\rm max} = 49/46~R_{\rm WD}$) be lower than the typical
ionization temperature of helium (log~$T$(K)$\simeq~3.95$).

\begin{figure}
\plotone{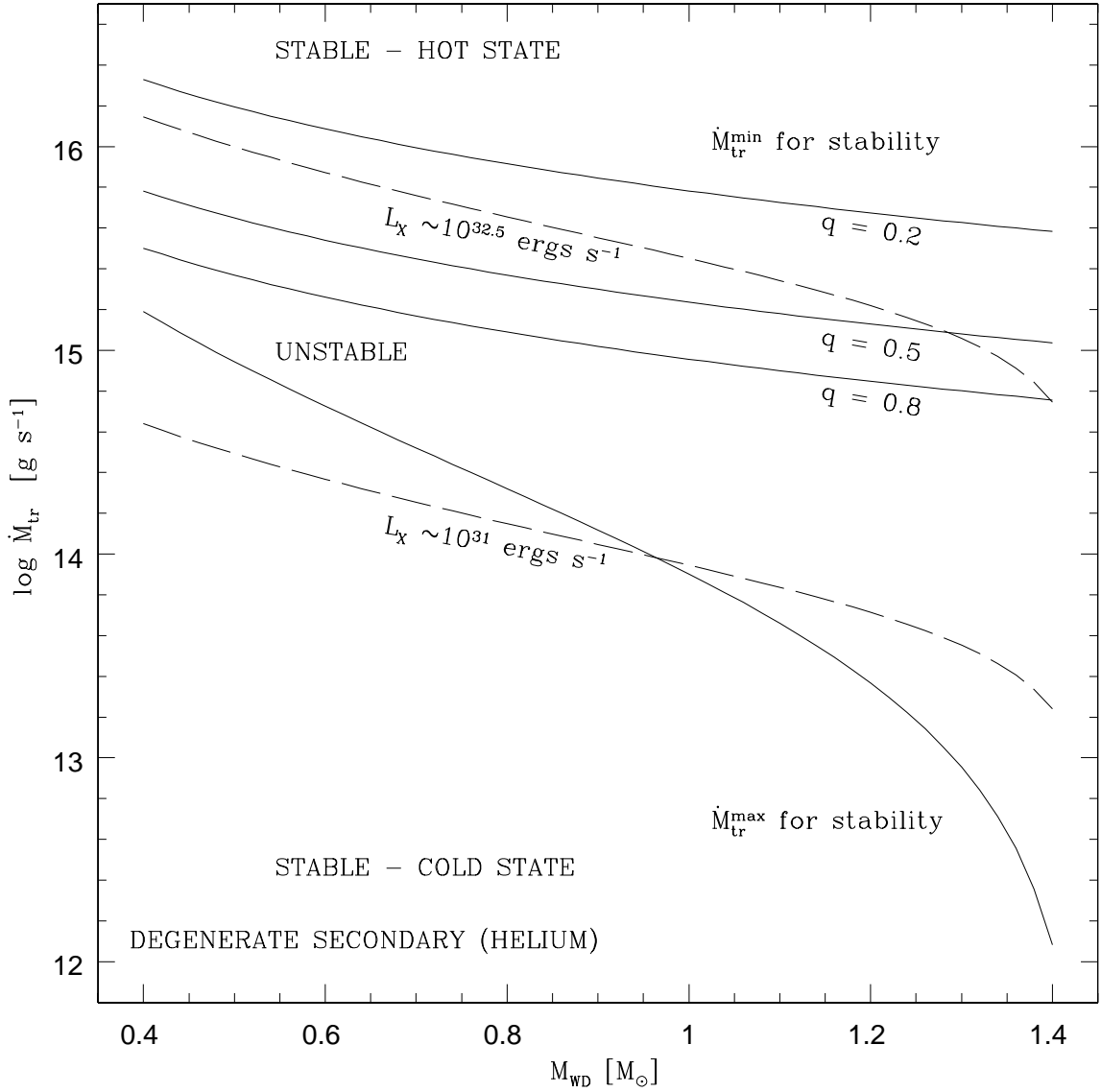}
\caption{Same as Fig.~\ref{fig:three} when the donor is assumed to be
a helium-rich degenerate star.}
\label{fig:five}
\end{figure}

Fig.~\ref{fig:five} shows the results of our stability analysis for AM
CVns in the usual format.  Although some of the candidate CVs in GCs
with X-ray luminosities at {the low end of the observed luminosity
function could be described as systems containing globally cold and
stable disks (depending on $M_{\rm WD}$), this does not appear to be a
solution for the entire population.  On the other hand, the faint
optical luminosities of candidate CVs in GCs also appear to be
inconsistent with globally hot and stable disks in these systems
\citep{Sharetal96,Ed03b}}

\subsection{Truncated disks}

A simple way of reconciling the observed X-ray
luminosities of candidate CVs in GCs with the DIM and the rare
occurrence of outbursts is to consider the possibility that disks in
most of these systems are sufficiently truncated to be stabilized. There is
growing evidence that even the so-called ``non-magnetic" CVs posses
magnetic moments sufficiently large ($\mu \gta 10^{30}\ \rm G\ cm^3$)
to partially truncate accretion disks in quiescent dwarf-novae \citep[see
e.g.][]{L04,ww,Warner04}.

\begin{figure}
\plotone{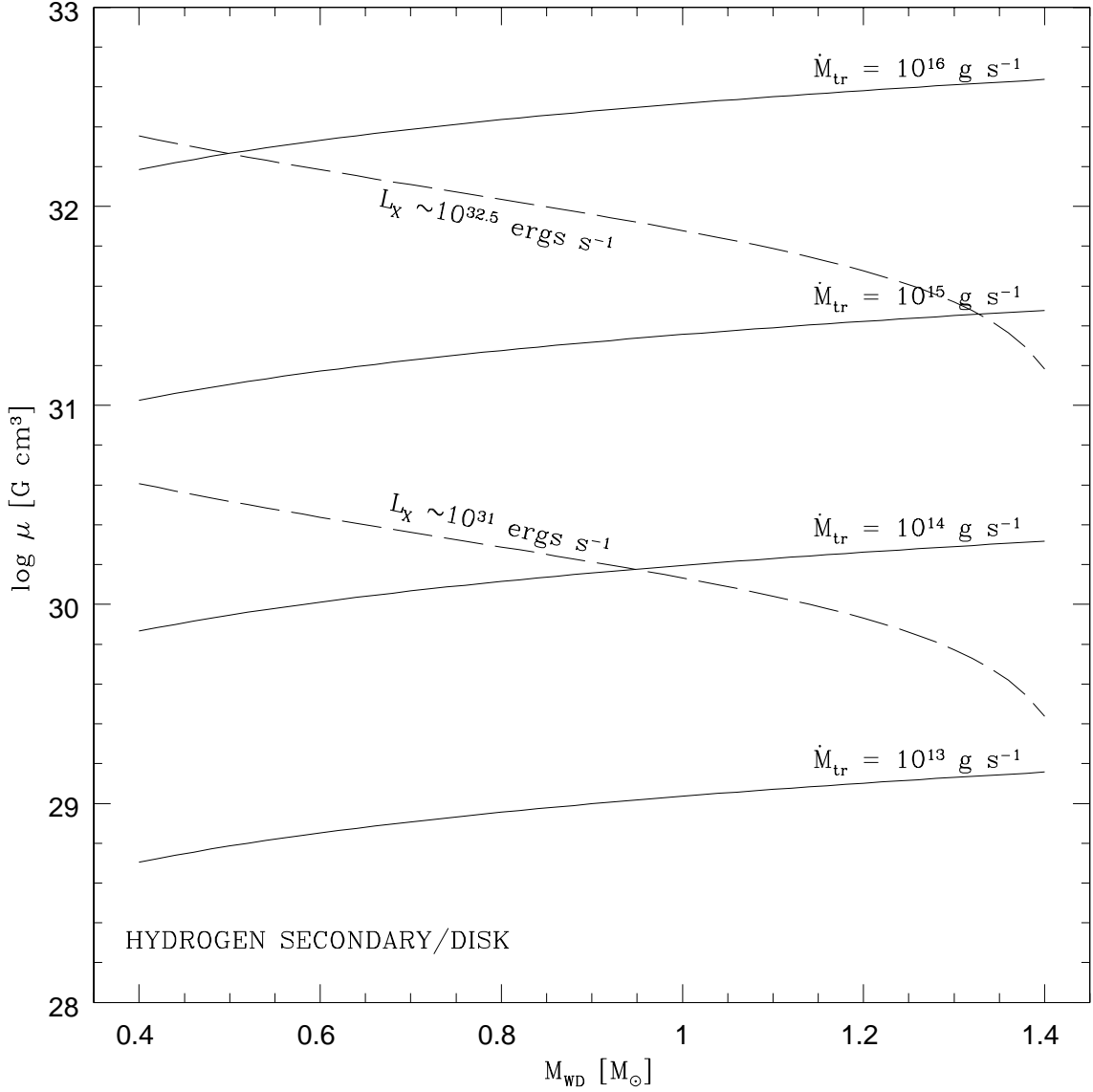}
\caption{Minimum values of the magnetic moment $\mu$ of an accreting
white-dwarf required to truncate the unstable regions of a
hydrogen-rich disk and stabilize it in the cold state, for four values
of the mass transfer rate, $\dot M_{\rm tr}$ (solid lines). White
dwarf masses in the range $M_{\rm WD}=0.4$-$1.4 M_\odot$ are
considered.  The mass transfer rate equivalents of $L_{\rm X} \sim
10^{31}$ and $10^{32.5}$~erg~s$^{-1}$, according to Eq.~(\ref{lx})
with $\eta_X=0.5$, are shown as long-dashed lines. Results for a
helium-rich disk would be similar except for a reduction in all
$\mu$--values by a factor $\sim 2$.}
\label{fig:six}
\end{figure}

Using the magnetospheric radius
\begin{equation}
R_{\rm m} = 9.8 \times 10^8 \left(\frac{\dot M}{10^{15}\rm g\ s^{-1}}
\right)^{-2/7} m_1^{-1/7} \left(\frac{\mu}{10^{30}\ \rm G\ cm^3}
\right)^{4/7} {\rm cm}
\end{equation}
and the stability criterion for a hydrogen disk in Eq. (\ref{mdmax1}),
one can deduce, for a given mass-transfer rate, the minimum value of
the magnetic moment required to truncate the unstable regions of the
disk and thus effectively stabilize the rest of it in the cold state:
\begin{equation}
\mu \geq 2.3 \times 10^{31} \left(\frac{\dot M}{10^{15}\rm g\
s^{-1}} \right)^{1.16} m_1^{0.831}\ \rm G~cm^3 \label{muc}
\end{equation}
Fig.~\ref{fig:six} shows critical values of the magnetic moments,
$\mu$, required to stabilize the disks for four different values of
the mass-transfer rate, $\dot{M}_{\rm tr} $ (solid
lines). Representative X-ray luminosities, calculated from
Eq.~(\ref{lx}) with $\eta_X=0.5$, are also shown for comparison
(dotted lines). Magnetic moments in the range $10^{29}$ -- $10^{32}$
G~cm$^3$ are sufficient to stabilize disks and at the same time be
consistent with the observed X-ray luminosities of candidate CVs in
GCs. Redoing the same analysis for disks composed of helium, we find
that the magnetic moments required to stabilize helium disks are only
slightly lower, by a factor $\sim 2$.

The uppermost values of the magnetic moments shown in
Fig~\ref{fig:six} correspond to magnetic moments of the intermediate
polar (IP) class but they would be required only for the highest X-ray
luminosities observed among candidate CVs in GCs. For the bulk of the
population, values of $\mu \gta 10^{30}\ \rm G\ cm^3$ are sufficient
to stabilize disks. These values may in fact be relevant to some of
the dwarf novae in the field. However, it is the combination of these
moderately strong magnetic moments with mass transfer rates lower than
in the field that is sufficient to stabilize the disks of candidate
CVs in GCs and explain their rare cataclysms.

\section{Discussion and Conclusion}

Using a series of quantitative stability criteria and detailed
time-dependent models, we have argued that it is not sufficient to
invoke lower mass transfer rates relative to CVs in the field to
explain the very infrequent outbursts of candidate CVs in GCs. The
argument relies on observed X-ray luminosities to set lower limits on
the rate of mass transfer in these systems. By contrast, we found that
a combination of moderately strong white dwarf magnetic moments and
low mass transfer rates can stabilize most disks in these systems and
therefore explain rare cataclysms in GCs.

Our proposed explanation for the rare occurrence of cataclysms in GCs
shares some similarities with a number of proposals previously made in
the literature \citep[e.g.][]{Ed03b,G99,Sharetal96,Sharetal04,ivanova}
but it is also specific in requiring that both magnetic truncation and
low mass transfer rates be present in these systems.  We have
calculated limits on the mass transfer rates and magnetic moments
which must be satisfied in candidate CVs in GCs for their disks to be
stabilized. As a result of these specific requirements, one should be
in a better position to put this theory to the observational test.

In that respect, we note that the comparatively low optical to X-ray
luminosity ratios of candidate CVs in GCs may already be providing
evidence in support of our scenario. At a given (low) mass transfer
rate (thus fixing the amount of X-ray produced in the vicinity of the
white dwarf), by truncating the inner, most optically luminous regions
of a CV accretion disk, one may indeed reduce the optical to X-ray
luminosity ratio of a system, relative to the untruncated
case. Detailed models would be required to address this point more
quantitatively.

Various properties of the few outbursts observed in GCs may also
contain valuable information on the regime of accretion present in
these systems, as we argued in \S\ref{sec:low} for fully extended
disks. Probably the most revealing observation for any candidate CV
in a GC would be a period and/or mass-transfer rate measurement, which
may tell us right away whether or not outbursts are expected from this
system, for a given value of the white dwarf magnetic moment. A
measurement of the magnetic field strength itself would obviously be a
most interesting test of our scenario, but this type of measurement is
expected to be very challenging for any GC given the field strength
values that we have been considering.

We have stressed throughout this work the need for, and the usefulness
of, robust statistical studies of outbursts in GCs. Given that some
outbursts may be missed by crowding effects in some clusters and that,
for instance, a count of the total number of CVs in a GC as
well-studied as 47 Tuc is barely known at the order-of-magnitude
level, it is a priori difficult to use existing reports on observed
outbursts to constrain the models in a meaningful way. \citet{Sha05}
have shown how completeness tests for a given cluster observational
campaign can greatly help interpret outburst searches in GCs: more
statistical studies of {this sort, combined with long-term monitoring
campaigns \citep[e.g.][]{Bond05,Kal05}, would clearly be useful.}

If our interpretation for the origin of rare cataclysms in globular
cluster CVs is correct, it points to lower average mass transfer rates
and stronger white dwarf magnetic moments in these systems than in the
field.  This would indicate important and systematic evolutionary
differences between field and cluster CV populations. Although such
differences may be expected on the basis of distinct dynamical
histories for CVs and their progenitors in GCs vs. the field
{\citep[e.g.][]{ivanova}}, our scenario is useful in making specific
predictions for what these differences might be. Mass transfer rates
$\dot M_T \gta 10^{14-15}$~g~s$^{-1}$ are comparable to or lower than
the already low value of $\sim 10^{15}$~g~s$^{-1}$ inferred for the
very long recurrence time system WZ Sge \citep{smak93}. These low
$\dot M_T$ requirements could thus provide important constraints on
the mechanism responsible for angular momentum loss in these binaries.
For instance, using the same $m_2$--$P_{\rm orb}$ relations as in
\S\ref{hdeg} and \S\ref{hedeg}, white dwarf masses in the range $0.4$
to $1.4 M_\odot$ and a standard formula for the rate of mass transfer
driven by gravitational wave braking \citep{Warner95}, one finds that
typical X-ray luminosities of candidate CVs in GCs can be explained by
orbital periods ranging from a fraction of an hour to several hours.

We note that, based on a study of white dwarf compressional heating,
\citet{towbil02} concluded that GCs may possess a rather large
population of CVs with very low mass-transfer rates ($\lta 6 \times
10^{14}$~g~s$^{-1}$), which could be post-turnaround binaries,
i.e. CVs which have evolved past the minimum orbital period.  Low mass
transfer rates also imply long accumulation times for classical nova
outbursts, which may thus be rarer in GCs than in the field. Finally,
the low values of mass transfer rates in CVs may be related to the
systematically low $\dot M_T$ values inferred by \citet{Heinke03} for
the population of quiescent LMXBs in GCs.

In the same vein, the typical values of white dwarf magnetic moments
required in our scenario are below IP-class values but they remain
larger than the values implicitly assumed in standard models of
erupting field dwarf novae (which are usually such that the disks are
never truncated, not even in quiescence; but see also
\citet{L04}). Systematically higher values of white dwarf magnetic
moments in GCs relative to the field, as suggested by our
interpretation of rare cataclysms, would then point to different
stellar/dynamical evolutionary histories for the progenitors of these
two CV populations \citep[e.g.][]{ivanova}.

\section*{Acknowledgments}

We thank Mike Shara for very useful and informative discussions {and
Craig Heinke for insightful comments on the manuscript, which helped
improve its quality.} A.D. acknowledges support from the Marie Curie
Fellowship program of the European Union under contract
No. HPMT-GH-00-00132-63 and the Slovak Academy of Sciences under Grant
No. 4015/4. This research was supported in part by the National
Science Foundation under Grant No. PHY99-07949 (at KITP).

\clearpage

\clearpage

\clearpage

\clearpage

\clearpage

\clearpage

\end{document}